\documentclass[11pt]{revtex4}
\usepackage{color}
\usepackage{amsmath,amssymb}
\usepackage[dvipdfm]{graphicx}
\usepackage{cancel}

\allowdisplaybreaks

\begin{document}
\title{Dark parameterization approach to Ito equation}

\author{Bo Ren$^{1}$\footnote{E-mail: renbosemail@gmail.com.}, Xi-Zhong Liu$^{1}$ and Ping Liu$^{2}$}

\affiliation{$^1$Institute of Nonlinear Science, Shaoxing University, Shaoxing, 312000, China \\
$^{2}$College of Electron and Information Engineering, University of Electronic Science and Technology
of China Zhongshan Institute, Zhongshan, 528402, China\\
e-mail: renbosemail@gmail.com}

\date{\today $\vphantom{\bigg|_{\bigg|}^|}$}

\begin{abstract}
The novel coupling Ito systems are obtained with the dark parameterization approach. By solving the coupling equations, the traveling wave solutions are constructed with the mapping and deformation method. Some novel types of exact solutions are constructed with the solutions and symmetries of
the usual Ito equation. In the meanwhile, the similarity reduction solutions of the model are also studied with the Lie point symmetry theory.
\end{abstract}

\maketitle


\section{Introduction}

Various astronomical and cosmological observations
show that there is dark matter (DM) and dark energy (DE) making up about
$20\%$ and $70\%$ of the energy budget of our universe respectively \cite{dark}. Although the evidence
for DM and DE has been established for many decades, the identity
of its basic constituents has so far remained elusive. Supersymmetry, a new symmetry that transforms bosons to fermions and vice versa in particle physics, still escapes observation \cite{super}.
A natural extension to the standard model
is the addition of a fourth generation of fermions with masses much larger than those of the three known generations \cite{fourg}.
These additional quarks can offer solutions to some outstanding theoretical questions, such as baryon asymmetry of the universe, Higgs naturalness and fermion mass hierarchy \cite{fourge}. However, up to now, no evidence for fourth generation quarks is observed at the Larger Hadron Collider.
To explain some kinds of indirectly detectable dark things, the dark parameters of the physical models have been proposed \cite{lousy}. Using this ideal to integrable systems, some novel coupled KdV and KP systems are obtained by the original usual field and partner fields. \cite{lousy,ling}. It is proved that the dark parameterization method can be successfully to find and solve new integrable systems, new localized excitation modes and new interaction phenomena among soliton or solitary waves.

In this letter, we shall use the dark parameterization method to the Ito equation. The usual Ito equation is
\begin{equation}\label{ito}
u_{tt} + 6 u_{xx} u_{t} + 6 u_{x} u_{xt} + u_{xxxt} = 0,
\end{equation}
which was first proposed by Ito, and its bilinear B\"{a}cklund transformation, Lax pair and multi-soliton solutions were obtained by means of Hirota's bilinear method \cite{Ito}. The properties of the equation such as the nonlinear superposition formula, Kac-Moody algebra, bi-Hamiltonian structure and supersymmetric version have been further proposed \cite{Huxb,liuphy,qpliu}.

The structure of this paper is organized as follows. In section 2, we present the general dark parameterization approach to integrable systems. The novel coupling Ito systems are given with the approach.
In section 3 and 4, we shall use one and two dark parameters as special examples.
The exact solutions of the coupling systems are found using the mapping and deformation method and Lie point symmetry.
The last section is a simple summary and discussion.

\section{General dark parameterization approach}

In this section, we consider the generalized dark parameterization approach to integrable system.
We assume that the real physical quantity $H$ include the usual quantity $H_0$ and the indirectly observed quantity $H_{i}$ \cite{lousy}
\begin{equation}\label{ham}
H=H_0+H_{i}(\alpha_1,\alpha_2,\cdots,\alpha_i)=H_0+H_{i}\alpha_i,
\end{equation}
where $\alpha_i$ are dark parameters and $H_0$ is independent of $\alpha_i$.
$H_i$ represents can not be directly observed quantities, such as DM, DE and supersymmetry. After introducing dark parameters into the traditional physical models, some types of partner fields may
be introduced.
With introducing this ideal to the nonlinear integrable systems, i.e., $\{H, H_i\} \rightarrow \{u, u_i\}$, one may obtain
infinitely many new coupled integrable systems which are constituted by the original usual field and partner fields.

The special type of the coupling Ito systems write as the following form combination the usual Ito equation \eqref{ito}
\begin{equation}\label{itoc}
\sum_{i=0}^{n}\Bigl(u_{i,tt} + u_{i,xxxt} + 6 \sum_{j=0}^i u_{{i-j},xx} u_{j,t} + 6 \sum_{j=0}^i u_{i-j,x} u_{j,xt}\Bigr)\alpha_i = 0.
\end{equation}
The function $u_0$ is exactly the usual Ito equation which has been widely studied \cite{Huxb,liuphy,qpliu}.
The partner fields $u_i (i\geq 1)$ are linear partial differential equations while the previous functions $u_j (j<i)$ are
known. We can theoretically solve the equivalent partner fields one after another.
Therefore, if we know the solution of $u_i$, then
we can construct a solution of the novel coupling Ito systems \eqref{itoc} via $u = u_i \alpha_i$.
In the next two sections, we discuss
two explicit examples for $n = 1$ and $n = 2$ respectively.

\section{One dark parameter Ito System and its solutions}

For $n=1$, \eqref{itoc} becomes
\begin{subequations}\label{itot}
\begin{align}
&u_{0,tt} + u_{0,xxxt} + 6 u_{0,xx} u_{0,t} + 6 u_{0,x} u_{0,xt}  = 0, \\
&u_{1,tt} + u_{1,xxxt} + 6 u_{1,xx} u_{0,t} + 6 u_{0,xx} u_{1,t} + 6 u_{1,x} u_{0,xt} + 6 u_{0,x} u_{1,xt} = 0.
\end{align}
\end{subequations}
The above system may has the different meanings with selecting different kinds of parameter $\alpha_1$.
If we simply take $\alpha_1$ as the Grassmann number, then $u_1$
is fermionic and the model is just the supersymmetric Ito version for fermionic component field \cite{qpliu}.
If we take the dark parameter $\alpha_1 = \zeta_1\zeta_2$ as a multiplication
of two Grassmann numbers, then both $u_0$ and $u_1$
are bosonic field \cite{ren}. The solution of the one dark parameter Ito system \eqref{itoc} can be constructed
via $u = u_0 + \alpha_1 u_1$ by means of \eqref{itot}.

\subsection{Traveling wave solutions with mapping and deformation method}

Introducing the traveling wave variable $X=k x+\omega t + c_0$ with constants $k$, $\omega$
and $c_0$, \eqref{itot} is transformed to the ordinary differential equations (ODEs) and directly integrate once
\begin{subequations}
\begin{align}\label{fourcom}
& k^3 u_{0,XXX} + \omega u_{0,X} + 6 k^2 u_{0,X}^2 = 0, \\
& k^3 u_{1,XXX} + \omega u_{1,X} + 12 k^2 u_{1,X} u_{0,X} = 0.
\end{align}
\end{subequations}
As the well known exact solutions of (5a), we try to build the mapping and deformation relationship between $u_{0}$ and $u_{1}$.
We get $u_{0,X}$ with (5a)
\begin{align}\label{u0}
u_{0,X} = -\frac{u_0^2}{k} - \frac{\omega}{4k^2}.
\end{align}
In order to get the mapping relationship between $u_1$ and $u_0$, we introduce the variable transformation
\begin{align}\label{trans}
u_1(X)=U_1(u_0(X)).
\end{align}
Using the transformation \eqref{trans} and vanishing $u_{0,X}$ via \eqref{u0}, the linear ODEs (5b) becomes
\begin{align}\label{map}
\bigl(4 k u_0^2 + \omega \bigr) \frac{\emph{d}^3 U_{1} }{\emph{d} u_0^3}  + 24 k u_0 \frac{\emph{d}^2 U_{1} }{\emph{d} u_0^2} - 24 k \frac{\emph{d}U_{1}}{\emph{d} u_0} =0.
\end{align}
The mapping and deformation relations are constructed via \eqref{map}
\begin{align}\label{mapp}
U_{1} = & c_1 + c_2 (20ku_0^2+\omega) + c_3\biggl( (20ku_0^2+\omega) \int^{u_0} \frac{6\sqrt{k\omega}}{(4k y^2 + \omega)} dy + \frac{2\sqrt{k\omega}u_0( 60ku_0^2+ 13\omega)}{4k u_0^2 + \omega} \biggr),
\end{align}
where $c_i \,(i=1,2,3)$ are arbitrary constants.
If we know the solution of $u_0$, the traveling wave solution $u_1$ will be given with considering \eqref{trans} and \eqref{mapp}.
Here, we list one solution as an example. The solution of $u_0$ can be expressed as the following form using \eqref{u0}
\begin{align}\label{uu0}
u_{0} = - \frac{1}{2}\sqrt{\frac{\omega}{k}} \tan \biggr(\frac{\sqrt{k\omega}(X+c_{4})}{2k^2} \biggl).
\end{align}
We can get the solution of the one dark parameter Ito system \eqref{itoc} with \eqref{mapp} and \eqref{uu0}
\begin{align}\label{solp}
u = & - \frac{1}{2}\sqrt{\frac{\omega}{k}} \tan \xi + \alpha_1 \Bigl[c_1 -  c_2 \omega \bigl(4 - 5 \sec^2 \xi\Bigr) + \\ \nonumber
& c_3 \omega \Bigl(\sin 2\xi - 15\tan\xi +12 \arctan(\tan\xi) -15\arctan(\tan\xi) \sec^2 \xi \Bigr) \Bigr],
\end{align}
where $\xi=\frac{\sqrt{k\omega}(X+c_{4})}{2k^2}$.

Besides, $u_{1}$ of (4b) exactly satisfies the
symmetry equation of the usual Ito system. For any given a solution $u_{0}$ of the usual Ito equation, a certain type solutions \eqref{itot} can be constructed
\begin{align}
u = u_0 + \alpha \sigma(u_{0}),
\end{align}
where $\sigma(u_{0})$ represents the symmetry of the usual Ito equation (4a).
It means that we have much freedom to choose $u_0$ so as to construct solutions of
the \eqref{itot}. The solution $u_0$ is not restricted to the traveling
wave solutions. All in all, we can construct not only traveling
wave solutions but also some novel types of solutions of
\eqref{itot} with the solutions and symmetries of the Ito equation.
As illustrative example, the $N$-soliton solution of the Ito equation reads \cite{Ito}
\begin{align}
u_{Ito} = 2 \biggl[ \log \Bigl( 1 + \sum^N_{k=1} \sum_{i_1>i_2>\cdots>i_k} \prod_{m>n} A_{i_mi_n} \exp\sum^k_{i=1} \eta_i \Bigr) \biggr]_{xx},
\end{align}
where $A_{ij} = \frac{(k_i-k_j)(k_i^3-k_j^3)}{(k_i+k_j)(k_i^3+k_j^3)}$ which is the phase shift from the interaction of the soliton $``i"$ with the soliton $``j"$, $\eta_i=k_ix-k^3_it+\eta_{i}^0$ and arbitrary constants $(k_i,\, \eta^0_i,\, \eta_i,\, i=1,2,\cdots,N)$.
Correspondingly, a special type of multiple soliton solutions of \eqref{itot} reads
\begin{align}
u = u_{Ito} + \alpha (x u_{Ito,x} + 3t u_{Ito,t} + u_{Ito} ).
\end{align}

\subsection{Similarity reduction solutions with symmetry reduction approach}

The symmetry study plays a prominent role in
nonlinear partial differential systems for the existence of infinitely many symmetries~\cite{ss7,ss8}. The classical
Lie group~\cite{ss4}, the nonclassical approach~\cite{ss5}, Clarkson and Kruskal (CK) direct method \cite{ss6}
are effective methods to obtain the
explicit exact solutions. Now, we shall use the Lie point symmetry approach to study \eqref{itot}.

A Lie point symmetry vector field is given
\begin{align}
V = X \frac{\partial}{\partial x} +  T \frac{\partial}{\partial t} + U_0 \frac{\partial}{\partial u_0} + U_{1} \frac{\partial}{\partial u_{1}},
\end{align}
where $X$, $T$, $U_0$ and $U_{1}$ are functions of $x$, $t$, $u_0$ and $u_{1}$. It means the system of \eqref{itot} is invariant under
\begin{align}
\{x, t, u_0, u_{1}\} \rightarrow \{x+\epsilon X, t+\epsilon T, u_0+\epsilon U_0, u_{1}+\epsilon U_{1}\},
\end{align}
with an infinitesimal parameter $\epsilon$. The corresponding symmetry can be supposed
\begin{align}\label{symm}
\sigma_0 = Xu_{0,x} + Tu_{0,t} - U_0, \,\, \,\, \,\, \sigma_{1} = Xu_{1,x} + Tu_{1,t} - U_{1}.
\end{align}
Considering the notation \eqref{symm}, $\sigma_{0;1}$ is the solution of the linearized \eqref{itot}
\begin{subequations}\label{fosym}
\begin{eqnarray}
& \sigma_{0,tt} + \sigma_{0,xxxt} + 6 \sigma_{0,t} u_{0,xx} + 6 \sigma_{0,xx} u_{0,t} + 6 (\sigma_{0,x} u_{0,x})_t = 0, \\
& \sigma_{1,tt} + \sigma_{1,xxxt} + 6 \sigma_{0,t} u_{1,xx} + 6 \sigma_{1,xx} u_{0,t} + 6 \sigma_{1,t} u_{0,xx} + 6 \sigma_{0,xx} u_{1,t} + 6 (\sigma_{0,x}u_{1,x})_t + 6(\sigma_{1,x}u_{0,x})_t = 0. \nonumber \\ &
\end{eqnarray}
\end{subequations}
Substituting \eqref{symm} into the symmetry equations \eqref{fosym} with $u_0$ and $u_{1}$ satisfying \eqref{itot}, we obtain the determining equations by identifying all coefficients of derivatives of $u_0$ and $u_{1}$. The solutions of the functions $X$, $T$, $U_0$ and $U_{1}$ can be concluded using the determining equations
\begin{align}\label{symso}
T = C_1 t + C_2, \hspace{0.6cm} X=\frac{C_1}{3}x + C_3, \hspace{0.6cm} U_0=-\frac{C_1}{3}u_0 + C_5,  \hspace{0.6cm} U_{1} = C_4 u_{1}+C_6,
\end{align}
where $C_i \,(i=1,2,...,6)$ are arbitrary constants.
Then, one can solve the characteristic equations to obtain similarity solutions
\begin{align}
\frac{\emph{d} x}{X}=\frac{\emph{d}t}{T}, \hspace{0.8cm} \frac{\emph{d} u_0}{U_0}=\frac{\emph{d}t}{T}, \hspace{0.8cm} \frac{\emph{d} u_{1}}{U_{1}}=\frac{\emph{d}t}{T}.
\end{align}

An simple example is listed concerning the solutions of \eqref{itot} in the following.
With $C_1=C_4=C_5=C_6=0$, we can find the similarity solutions after solving out the characteristic equations
\begin{align}\label{slotu}
u_0=U_0(\xi), \hspace{1.2cm}u_{1}=U_{1}(\xi),
\end{align}
with the similarity variable $\xi=t-\bigl(C_2/C_3\bigr)x$. We redefine the similarity variable as $\xi=x+ct$ with $c$ an arbitrary velocity constant. Substituting \eqref{slotu} into \eqref{fosym}, the invariant functions $U_0$ and $U_1$ satisfy the reduction systems
\begin{subequations}\label{similarity}
\begin{align}
&U_{0,\xi\xi\xi\xi} + c U_{0,\xi\xi} + 12 U_{0,\xi\xi} U_{0,\xi} =0,\\
&U_{1,\xi\xi\xi\xi} + c U_{1,\xi\xi} + 12 (U_{1,\xi} U_{0,\xi})_\xi =0.
\end{align}
\end{subequations}
These reduction equations are linear ODEs while the previous functions are known, we can theoretically solve \eqref{similarity} one after another.

\section{Two dark parameters Ito System and its solutions}

For $n=2$, \eqref{itoc} becomes
\begin{subequations}\label{itoth}
\begin{align}
&u_{0,tt} + u_{0,xxxt} + 6 u_{0,xx} u_{0,t} + 6 u_{0,x} u_{0,xt}  = 0, \\
&u_{1,tt} + u_{1,xxxt} + 6 u_{1,xx} u_{0,t} + 6 u_{0,xx} u_{1,t} + 6 u_{1,x} u_{0,xt} + 6 u_{0,x} u_{1,xt} = 0,\\
&u_{2,tt} + u_{2,xxxt} + 6 u_{2,xx} u_{0,t} + 6 u_{0,xx} u_{2,t} + 6 (u_{1,x} u_{1,t})_x + 6 (u_{2,x} u_{0,x})_t = 0.
\end{align}
\end{subequations}
The solution of the two dark parameters Ito system \eqref{itoc} can be constructed
via $u = u_0 + \alpha_1 u_1+ \alpha_2 u_2$ with solving \eqref{itoth}.

\subsection{Traveling wave solutions with mapping and deformation method}

With the variable $X=k x+\omega t + c_0$, \eqref{itoth} is transformed to the ODEs and integrate once
\begin{subequations}
\begin{align}
& k^3 u_{0,XXX} + \omega u_{0,X} + 6 k^2 u_{0,X} u_{0,X}= 0, \\
& k^3 u_{1,XXX} + \omega u_{1,X} + 12 k^2 u_{1,X} u_{0,X} = 0,\\
& k^3 u_{2,XXX} + \omega u_{2,X} + 12 k^2 u_{2,X} u_{0,X} + 6 k^2 u_{1,X}^2 = 0.
\end{align}
\end{subequations}
We consider the variable transformations
\begin{align}\label{transfo}
u_1(X)=U_1(u_0(X)), \hspace{1cm} u_2(X)=U_2(u_0(X)).
\end{align}
With the above transformations and eliminating $u_{0,X}$ via \eqref{u0}, the ODEs (24b) and (24c) are changed
\begin{subequations}\label{solth}
\begin{align}
&\bigl(4 k u_0^2 + \omega \bigr) \frac{\emph{d}^3 U_{1} }{\emph{d} u_0^3} + 24k u_0 \frac{\emph{d}^2 U_{1} }{\emph{d} u_0^2} - 24 k \frac{\emph{d} U_{1} }{\emph{d} u_0} =0, \\
&\bigl(4 k u_0^2 + \omega \bigr) \frac{\emph{d}^3 U_{2} }{\emph{d} u_0^3} + 24k u_0 \frac{\emph{d}^2 U_{2} }{\emph{d} u_0^2} - 24 k \frac{\emph{d} U_{2} }{\emph{d} u_0} - 24 k \Bigl(\frac{\emph{d} U_{1} }{\emph{d} u_0}\Bigr)^2 =0.
\end{align}
\end{subequations}
By repeating the processes of the last section, the traveling wave solution can be obtained using \eqref{transfo} and \eqref{solth}. The expression solution of $U_2$ is very length so that we neglect write it.

\subsection{Similarity reduction solutions with symmetry reduction approach}

For the two dark parameters Ito system \eqref{itoth}, the solutions of the functions $X$, $T$, $U_0$, $U_1$ and $U_{2}$ read
\begin{align}\label{symsol}
&T = C_1 t + C_2, \hspace{0.6cm} X=\frac{C_1}{3}x + C_3, \hspace{0.6cm} U_0=-\frac{C_1}{3}u_0 + C_4,  \hspace{0.6cm} U_{1} = \frac{3C_5-C_1}{6} u_{1}+C_6, \nonumber \\ & U_2=C_5u_2 +C_7u_1 + C_8,
\end{align}
where $C_i \,(i=1,2,...,8)$ are arbitrary constants. The similarity solutions can be
obtained by solving the characteristic equations.

An explicit example is discussed in the following.
With $C_1=C_4=C_5=C_6=C_7=C_8=0$, the invariant solutions are given
\begin{align}\label{slot}
u_0=U_0(\xi), \hspace{0.9cm} u_{1}=U_{1}(\xi),\hspace{0.9cm} u_{2}=U_{2}(\xi).
\end{align}
The reduction equations lead to
\begin{subequations}\label{tsimila}
\begin{align}
&U_{0,\xi\xi\xi\xi} + c U_{0,\xi\xi} + 12 U_{0,\xi\xi} U_{0,\xi} =0,\\
&U_{1,\xi\xi\xi\xi} + c U_{1,\xi\xi} + 12 (U_{1,\xi} U_{0,\xi})_\xi =0,\\
&U_{2,\xi\xi\xi\xi} + c U_{2,\xi\xi} + 12 (U_{0,\xi} U_{0,\xi})_\xi + 12 U_{1,\xi\xi} U_{1,\xi}=0,
\end{align}
\end{subequations}
where the similarity variable $\xi=x+ct$.
Similar to the one dark parameter Ito system, \eqref{tsimila} can theoretically be solved one after another.

\section{Conclusions}

In this paper, the coupled Ito systems are obtained with the dark parameterization approach. Using the the mapping and deformation method, the traveling wave solutions of the coupled systems are obtained. Besides, some special types of exact solutions can be given straightforwardly
through the exact solutions of the Ito equation and its symmetries. In addition, the similarity reduction solutions of the model are derived using the Lie point symmetry theory. The dark parameterization procedure can be applicable all the models including nonlinear integrable or non-integrable systems. The method is worthy of further studying.

\begin{acknowledgments}
This work was supported by the National Natural Science Foundation
of China (Grant Nos. 11305106 and 11305031), the Natural Science Foundation of Zhejiang Province
(Grant No. LQ13A050001) and the Natural Science Foundation of Guangdong Province (No. S2013010011546).
\end{acknowledgments}


\end{document}